# Optically Detected Cross-Relaxation Spectroscopy of Electron Spins in Diamond

Hai-Jing Wang, Chang S. Shin, Scott J. Seltzer, Claudia E. Avalos, Alexander Pines, and

Vikram S. Bajaj[*]

*Materials Sciences Division, Lawrence Berkeley National Laboratory, Berkeley,*

*California 94720, USA*

*Department of Chemistry and California Institute for Quantitative Biosciences,*

*University of California, Berkeley, California 94720, USA*

**The application of magnetic resonance (MR) spectroscopy at progressively smaller length scales may eventually permit "chemical imaging" of spins at the surfaces of materials and biological complexes. In particular, the negatively charged nitrogen-vacancy (NV⁻) centre in diamond has been exploited as an optical transducer for nanoscale nuclear magnetic resonance. However, the spectra of detected spins are generally broadened by their interaction with proximate paramagnetic NV⁻ centres through coherent and incoherent mechanisms. Here we demonstrate a detection technique that can resolve the spectra of electron spins coupled to NV⁻ centres, namely substitutional nitrogen ($N_S$) and neutral nitrogen-vacancy ($NV^0$) centres in diamond, through optically detected cross-relaxation. The hyperfine spectra of these spins are a unique chemical identifier, suggesting the possibility, in combination with recent results in diamonds harbouring shallow NV⁻ implants, that the spectra of spins external to the diamond can be similarly detected.**

The NV⁻ centre in diamond has been exploited in optically detected magnetic resonance (ODMR) experiments because of its favourable properties including spin-dependent fluorescence[1], spin coupling to the magnetic environment[2], and its long polarization lifetime as compared to other similar substrates[3], even at room temperature. Recent experiments have employed single NV⁻ centres to detect distant nuclear spins in diamond[4-8] and external, nanoscale nuclear spin ensembles[9,10], laying the foundations for routine MR imaging with nanometre-resolution[11,12]. Electron spins, in the form of either paramagnetic centres in the diamond lattice or radicals intimately associated with diamond surface, have also been optically detected through their coupling to the NV⁻ centre[13-15]. The characteristic spectra of the detected spins, however, are often broadened by these same interactions[13-15]. The spectroscopic details are, in general, not sufficiently resolved to obtain critical chemical information.

One demonstration of optically detected spectroscopy of the electron spin using the NV⁻ centre was performed by sweeping the magnetic field through the cross-relaxation condition between the NV⁻ and the substitutional nitrogen defect ($N_S$) near ~51.4 mT, where the two electron spin species have commensurate transition frequencies[1] (Fig. 1). However, this method requires resonant matching at a specific magnetic field such that spin flip-flop (zero-quantum) processes become allowed; this prevents it from serving as a generic method for optical detection of dark electron or nuclear spins. Alternatively, cross-relaxation can also occur in a weak magnetic field (< 5 mT, Fig. 1e), as previously suggested by observed dips in the NV⁻ spin-lattice relaxation time ($T_1$) and in the intensity of its zero-phonon line in this regime[16-19]. Here we instead demonstrate optically detected spectroscopy of electron spins, namely the $N_S$ and $NV^0$ centres in diamond, by identifying their respective spectral characteristics in the NV$^-$ cross-relaxation ODMR spectrum. This technique is the only ODMR method that can provide detailed spectra of the detected electron spins. Such information is essential for the detection and identification of radicals or relaxation centres by the NV$^-$ centre, and for the transduction of chemically informative NMR spectra by optical means.

We demonstrate this technique on an ensemble of NV$^-$ centres in a single-crystal diamond, host to several types of paramagnetic defects that may be identified from the ODMR spectra as shown in Fig. 2 (see Methods). The measured spectra have chemically informative fine features that differ vastly from ODMR spectra based on the allowed electron spin transitions of the NV$^-$ centre alone[20,21]. The majority of the fine features can be assigned to the $N_S$ centre[16,17,19,22] (see Supplementary Information), from which almost all calculated transitions (blue lines on top of each spectrum) are present in the observed spectra. For instance, the three calculated transition frequencies at $B_z \sim 0$ mT (i.e., ambient field), namely 18.4, 130.2, and 148.6 MHz (Fig. S1a), are matched with peaks in the ODMR spectrum (Fig. 2a). The intensity of the peaks agrees well with the 1:3 population ratio of $N_S$ centres oriented either parallel to $B_z$ (for example, the peak at 240 MHz in Fig. 2d) or ~109° relative to $B_z$ (the peak at 223 MHz in Fig. 2d). Although the $N_S$ centre is a dark spin whose hyperfine parameters were determined by electron paramagnetic resonance (EPR)[22], our observations show that the characteristic spectra of the $N_S$ centres can also be resolved by ODMR.

With the majority of the fine spectral features assigned to the $N_S$ centre, the remaining peaks are consistent with transitions from the NV$^0$ centre in the spin-3/2 $^4A_2$ excited state[23-25] (purple lines on top of each spectrum, see Supplementary Information).

There is excellent agreement between calculations and the observed transitions at 60, 70, 84, and 92 MHz at 1.44 mT (Fig. 2b), demonstrating that the $NV^0$ centre is optically detected with part of its characteristic spectrum resolved. As the magnetic field increases above 2 mT, most transitions of the $NV^0$ centre overlap with those of the $N_S$ centre, suggesting that weak magnetic field is required to separately resolve the two species. Although both $NV^-$ and $NV^0$ centres are photoluminescent[26], the $NV^0$ centre has not been detected previously using ODMR (unlike the $NV^-$ centre), and EPR spectra of the $NV^0$ centre have been observed only recently for its $^4A_2$ exited state under laser illumination[23]. We demonstrate here that the $^4A_2$ exited state of the $NV^0$ centre can also be detected by ODMR.

Cross-relaxation provides a generic mechanism to convert the magnetic resonance of $N_S$ and $NV^0$ spins into changes in the steady-state fluorescence intensity of the $NV^-$ centre. At weak magnetic field, $NV^-$-$N_S$ or $NV^-$- $NV^0$ spin flip-flop processes become possible because the transition energy can be matched through contact with the multibody dipolar degrees of freedom of the system. These flip-flop processes can also exchange angular momentum with lattice, and multiple-quantum spin-interaction elements, such as $S_+^i S_+^j$ and $S_-^i S_-^j$ (where $i$ and $j$ denote different types of paramagnetic centres), could have non-negligible contributions to the cross-relaxation[27]. Thus, when resonant RF saturates one of the $N_S$ or $NV^0$ spin transitions, cross-saturation then leads to enhanced spin-lattice relaxation of the $NV^-$ electron spin, and thus decreases the polarization of the "brighter" $|0\rangle$ sublevels, leading to decreased fluorescence intensity[1].

In addition to the fine features in the ODMR spectra, relatively broad peaks with full-width-at-half-maximum (FWHM) of ~20 MHz are also observed. The central frequency of each broad peak agrees well with the calculated energy difference between magnetically inequivalent NV⁻ centres that are oriented either parallel to $B_z$ or ~109° relative to $B_z$. The calculated energy difference is labelled by asterisks in Fig. 2 (see the Supplementary Information). For instance, the broad peak at ~70 MHz in Fig. 2d matches the energy difference between the $|+1\rangle$ sublevel of NV⁻ parallel to $B_z$ and the $|-1\rangle$ level of NV⁻ ~109° relative to $B_z$. Here, the RF enhances cross-relaxation by providing the energy difference between two NV⁻ centres, which otherwise need to exchange energy via phonons[28].

Cross-relaxation between magnetically inequivalent NV⁻ centres can be directly observed by measuring the spin-lattice relaxation time of the NV⁻ centre parallel to $B_z$ while saturating the transition of the NV⁻ centre ~109° relative to $B_z$ (see Methods). Figure 3 shows the apparent spin-lattice relaxation time ($T_1^*$), measured at the transition frequency of the NV⁻ centres oriented parallel to $B_z$, as a function of the saturation microwave (MW) frequency. The relaxation time becomes much shorter when the saturation frequency is on resonance with the allowed transitions of the NV⁻ centres ~109° relative to $B_z$. Note that saturation of dark electron spin transitions does not cause an observable change in the $T_1^*$ of the NV⁻ centre because of the relatively low polarization of the $N_S$ and NV⁰ centres.

Cross-relaxation spectroscopy can also be observed near the ground-state level anti-crossing (GSLAC) at ~102.5 mT, where the $|0\rangle$ and $|-1\rangle$ sublevels of the NV⁻ centre approach degeneracy and the $|0\rangle \leftrightarrow |+1\rangle$ transition frequency of the NV⁻ centre is almost twice the transition frequencies of the $N_S$ and NV⁰ centres (Fig. 1e). This cross-relaxation can therefore occur through a NV⁻ flip paired with the flops of two electron spins with the balance of the angular momentum transferred to the rigid lattice[27]. The characteristic peaks of the $N_S$ and NV⁰ centres are clearly resolved and identified in the ODMR spectrum within the 2.7-3.0 GHz range (Fig. 4). Four possible orientations of the $N_S$ centres lead to two triplets (blue bars on the bottom of the spectrum in Fig. 4) with overlapping central peaks when the external field is aligned with one of the $N_S$ (or NV⁻) symmetry axes[1,15,29]. The remaining peaks agree well with the calculated $|m_{S^0} = -1/2 \leftrightarrow +1/2\rangle$ transition frequencies of NV⁰ centres oriented either parallel to $B_z$ or ~109° relative to $B_z$, with offsets may due to the accuracy of the available spin Hamiltonian parameters that was determined for the ¹⁵NV⁰ system at the spin-3/2 $^4A_2$ excited state[23]. This transition has not been observed by EPR, probably because it is obscured by other paramagnetic centres in diamond[23]. Near GSLAC, the nearly complete electron spin polarisation of the NV⁻ centres can be transferred to the neighbouring electron spins via their mutual interactions, resulting in greater polarization[29] and thus signals from the $N_S$ and NV⁰ electron spins, similar to observations at the cross-relaxation condition near ~51.4 mT[1].

Our cross-relaxation spectroscopy technique generically detects the spectra of electron spins using the NV⁻ centre (see the illustration in Fig. S2). The steady-state fluorescence intensity of the NV⁻ centre is directly observed without requiring the creation or manipulation of NV⁻ spin coherences as in other methods. This effectively moves the detection of electron spins from the indirect to the direct frequency dimension, thereby improving both frequency resolution and detection sensitivity. Detection is no longer limited by the coherence time of the $NV^-$ centre, and as a result much longer detection times can be used in order to improve the signal-to-noise ratio. The fine frequency resolution and high sensitivity of this technique may also allow the identification of additional unknown paramagnetic centres that have been obscured by the $N_S$ centre in EPR[23], in a similar manner to our detection of the $NV^0$ centre. These results provide an optically detected analogue of nuclear magnetic cross-relaxation spectroscopy in solid and dilute spin systems[30,31].

It is also possible to use cross-relaxation spectroscopy for detection of electron spins external to the diamond sample. The potential detection distance through cross relaxation should be at least ~8 nm, as estimated by the average separation of neighbouring $NV^-$ centres in our diamond sample, between which cross relaxation is clearly detected in our measurements[32] (see Supplementary Information). Such a detection distance would allow for observation and identification of dark electron spins deposited on the diamond surface[14], if using shallow-implanted $NV^-$ centres in ultrapure diamond[3,33]. Furthermore, radicals on the diamond surface could be used as a detection intermediate for external nuclear spins located at even further distances[34]; an ensemble of radicals of a single type could significantly increase the detected magnetic flux[35], while spatially engineered arrays of radicals of different species could provide spatial resolution for detection of distant nuclear spins or simply act as a gradiometer[15,36].

## Methods

**Materials and experiments.** The diamond sample used in our experiments was fabricated commercially by Element-6 using high-pressure high-temperature (HPHT) synthesis. It was then irradiated, annealed, and characterized as previously reported[20,36], where it was labelled as sample S9. The diamond sample contains ~2 ppm of NV$^-$ and ~50 ppm of $N_S$ centres[20]. The average distance between nearest NV$^-$ centres is ~8 nm, and that between an NV$^-$ centre and the nearest $N_S$ centre is ~3 nm, as estimated by the Poisson distribution for the typical distance between neighbouring particles[37]. The ODMR apparatus was described in detail elsewhere[21] and we only briefly describe it here. A continuous beam from a 532 nm laser with optical power of 2 mW is focused on the diamond by an objective lens with a numerical aperture of 0.7 to achieve optical power density of ~ 5 mW/$\mu$m$^2$. The fluorescence signal from the NV$^-$ is collected by an avalanche photo detector after passing a dichroic mirror and a long-pass filter[36]. The external magnetic field is applied using a permanent magnet aligned to one of the NV$^-$ symmetry axes, with the field strength adjusted by the magnet's distance to the diamond and determined using the NV$^-$ centres as a magnetometer[20,21]. The radiofrequency (RF) radiation is swept from 1 to 300 MHz using a single copper wire loop to produce an oscillating magnetic field of ~0.1 mT in amplitude. The ODMR continuous-wave (CW) spectra record the steady-state fluorescence intensity under conditions of continuous RF excitation, compared to that with RF turned off for reference. In the double resonance experiments, pulsed microwave excitation is delivered using the same copper wire loop, while continuous microwave radiation is applied with an additional copper wire to saturate additional transitions. A standard inversion recovery sequence is used to measure the change in $T_1^*$ at the $|0\rangle \leftrightarrow |-1\rangle$ transition frequency of the NV⁻ centres oriented parallel to $B_z$ while saturating transitions of the NV⁻ centres oriented ~109° relative to $B_z$ at a different microwave frequency.

**Acknowledgements**

This work was supported by the Director, Office of Science, Office of Basic Energy Sciences, Materials Sciences and Engineering Division, of the U.S. Department of Energy under Contract No. DE-AC02-05CH11231.

**Author contributions**

H.J.W. and V.S.B. conceived the idea and planned the project, H.J.W., C.S.S., and C.E.A. constructed the experimental apparatus, H.J.W. and C.S.S. performed the experiments, H.J.W. and S.J.S. analysed the data, H.J.W., V.S.B., and S.J.S. wrote the manuscript, A.P. and V.S.B. oversaw the project, and all authors discussed the results and commented on the manuscript.

**Additional information**

Supplementary information is available in the online version of the paper.

**Competing financial interests**

The authors declare no competing financial interests.

**Figure Legends**

**Figure 1 | Transition frequency of each electron spin system in diamond as a function of the magnetic field.** The magnetic field is aligned along one of the NV⁻ axes (solid and open black squares: the NV⁻ centre parallel to and ~109° relative to $B_z$, respectively; solid blue triangles: the $N_S$ centre; solid and open purple circles: the NV⁰ centre parallel to and ~109° relative to $B_z$, respectively). The conditions where cross-relaxation occurs are labelled by green circles at $B_z = 0,\ 51.4,\ 59.5,\ 102.5\ \text{mT}$.

**Figure 2 | ODMR spectra acquired at weak magnetic field.** The spectra are acquired at (a) $B_z \approx 0\ \text{mT}$ (ambient field), (b) $B_z = 1.44\ \text{mT}$, (c) $B_z = 2.29\ \text{mT}$, and (d) $B_z = 3.89\ \text{mT}$. The fine features of the observed spectra are assigned to the $N_S$ and NV⁰ centres as coloured in blue and purple, respectively. Such assignment is based on the calculated transition frequencies of the $N_S$ (blue lines on the top of each spectrum) and NV⁰ (purple lines on the top of each spectrum) centres. All transitions from the $N_S$ centre and the relevant transitions of the NV⁰ centre oriented parallel to $B_z$ are shown. The short and long blue lines with 1:3 ratio in length correspond to the $N_S$ centres oriented parallel to $B_z$ and ~109° relative to $B_z$, respectively, with 1:3 ratio in population. The red asterisks (*) represents the difference in transition frequencies between the magnetically inequivalent NV⁻ centres.

**Figure 3 | Cross-saturation among magnetically inequivalent NV⁻ centres.** Cross-saturation between the NV⁻ centres oriented either parallel to $B_z$ or ~109° relative to $B_z$ at $B_z = 1.29\ \text{mT}$. The solid lines show the ODMR spectrum (left and bottom black axes) acquired near ~2.87 GHz, where the allowed transitions of NV⁻ centres at different

orientations can be resolved. The effective longitudinal relaxation time $T_1^*$ (scatter data with error bars representing the standard deviation, right and top blue axes) is measured at 2.834 GHz, the frequency of the $|0\rangle \leftrightarrow |-1\rangle$ transition of the NV⁻ centres oriented parallel to $B_z$, while saturating another MW frequency. The change in $T_1^*$ can be correlated very well with the ODMR spectrum; $T_1^*$ is significantly reduced when the saturating frequency is resonant with the transitions of the NV⁻ centres oriented ~109° relative to $B_z$. The solid and dashed horizontal lines are the average value and standard deviation of $T_1^*$ with the saturation MW turned off, respectively.

**Figure 4 | ODMR spectrum near GSLAC.** The ODMR spectrum is acquired at $B_z = 101.4$ mT. The most intense peak, centred at ~2.857 GHz and coloured orange, is an artefact whose second harmonic matches the frequency of the $|0\rangle \leftrightarrow |+1\rangle$ ground-state transition of the NV⁻ centre parallel to $B_z$. Each resolved side peak is fitted into a Gaussian function and assigned to either the $N_S$ centre (blue curves) or the NV⁰ centre (purple curve). The assignment is based on the calculated transition frequencies of each centre and its respective Hamiltonian, as shown by the vertical bars at the bottom. All allowed transitions of the $N_S$ centre in the ground state are clearly resolved at the predicted transition frequencies (blue bars). The rest of observed side peaks are consistent with the $|m_{S^0} = -1/2 \leftrightarrow +1/2\rangle$ transitions of the NV⁰ centre in the $^4A_2$ exited state with an offset from the predicted transition frequencies (purple bars) probably due to the accuracy of the available spin Hamiltonian parameters.

**Figure 1**

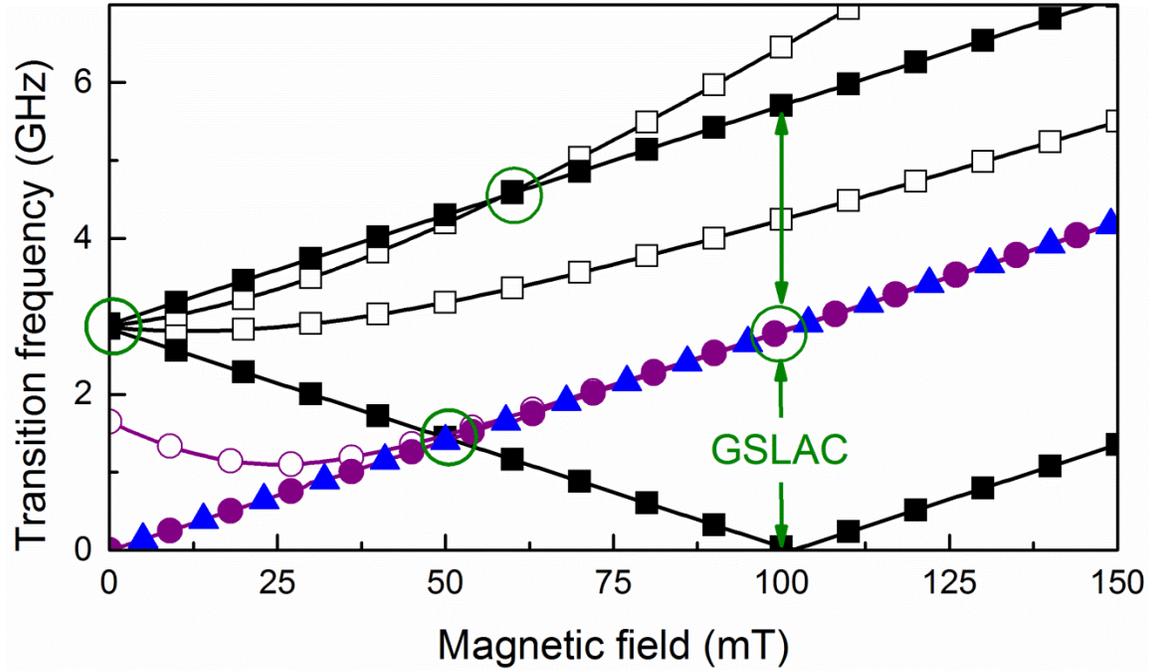

**Figure 2**

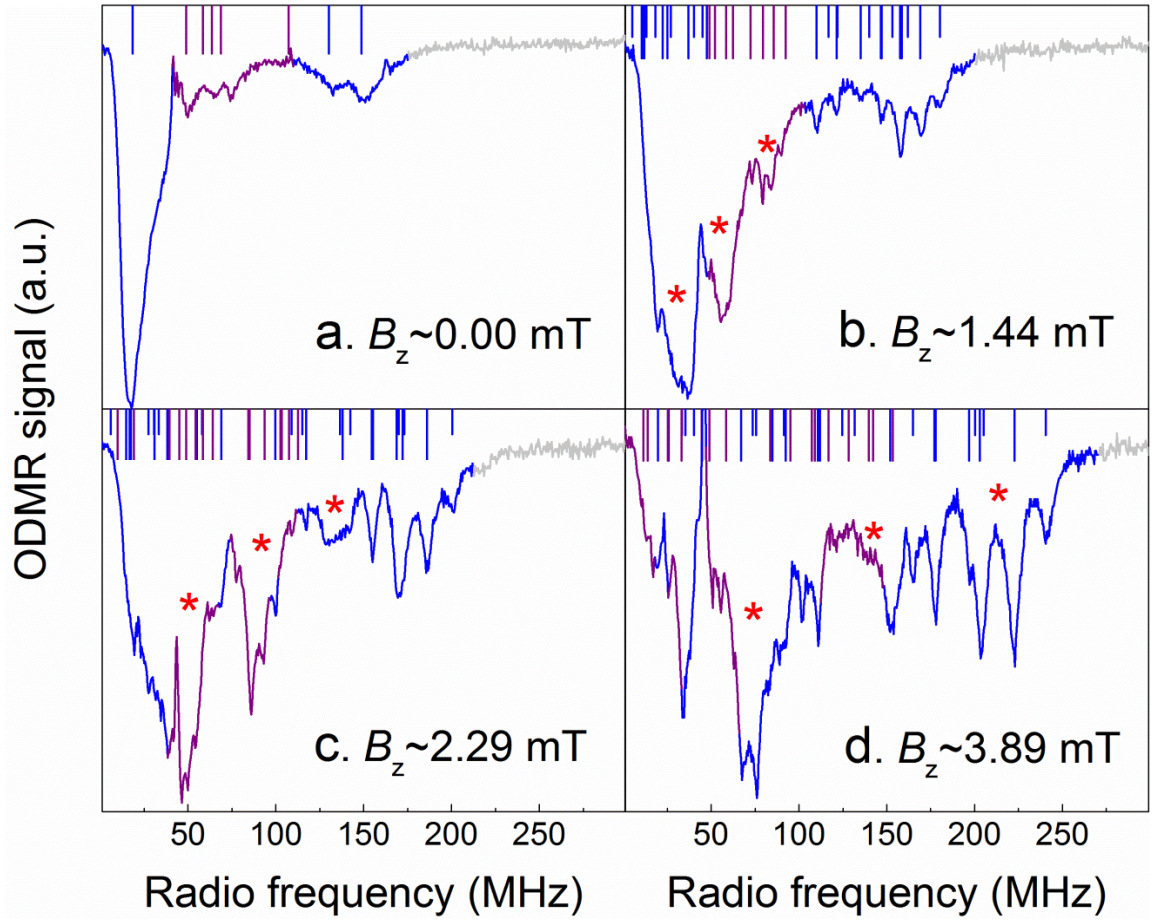

**Figure 3**

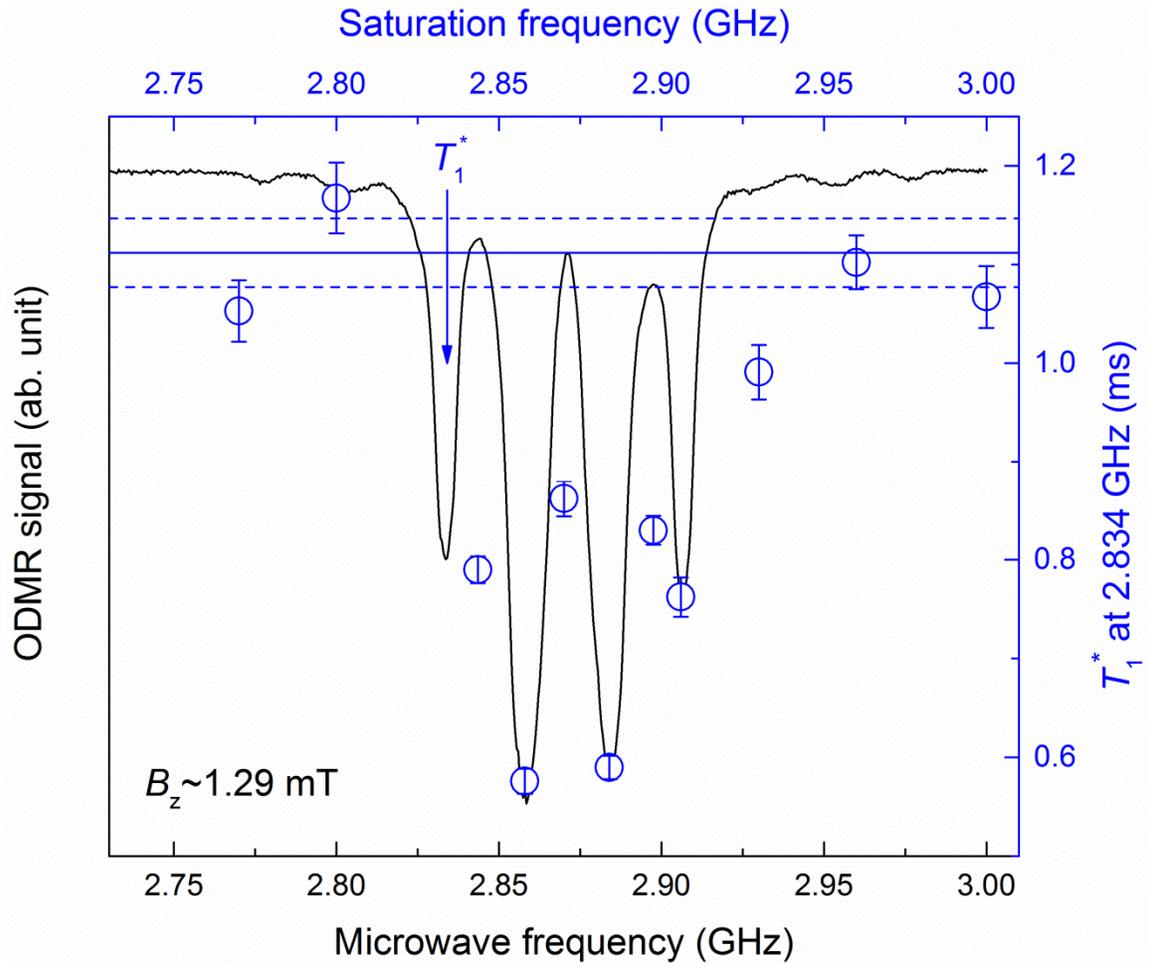

**Figure 4**

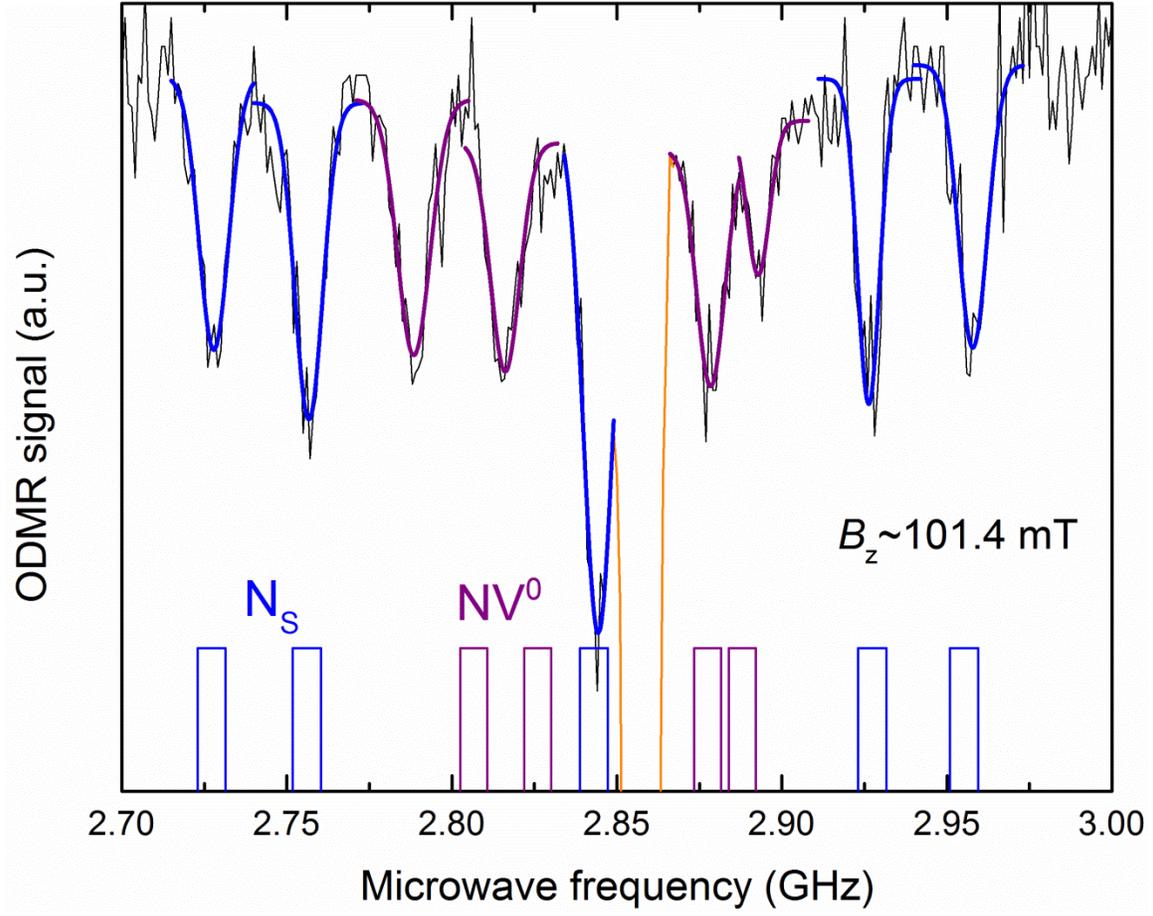

# Supplementary Information

# Optically Detected Cross-Relaxation Spectroscopy of Electron Spins in Diamond

Hai-Jing Wang, Chang S. Shin, Scott J. Seltzer, Claudia E. Avalos, Alexander Pines, and Vikram S. Bajaj.[*]

*Materials Sciences Division, Lawrence Berkeley National Laboratory, Berkeley, California 94720, USA*

*Department of Chemistry and California Institute for Quantitative Biosciences, University of California, Berkeley, California 94720, USA*

## 1 The Hamiltonian and calculated energy levels of each spin system

### 1.1 The NV⁻ centre

The NV⁻ centre has $C_{3v}$ symmetry and thus has four possible orientations within the diamond lattice (Fig. S1a). The NV⁻ electron spin forms a spin-1 triplet ($^3A$) in the ground state. The Hamiltonian is given approximately by[1,2]:

$$H_{NV^-} = D \cdot S_z^2 + E \cdot (S_x^2 - S_y^2) + \gamma_e \mathbf{B} \cdot \mathbf{S}, \qquad (S1)$$

where the axial ($D \sim 2.87$ GHz) and transverse ($E$, a strain splitting particular to each diamond sample; for our sample, $E \sim 3.5$ MHz) zero-field splitting (ZFS) parameters are determined from the spectra at ambient magnetic field[3], $\mathbf{S}$ is the electron spin of the NV⁻ centre with the electron gyromagnetic ratio of $\gamma_e$, $\mathbf{B}$ is the external magnetic field aligned along one of the NV⁻ symmetry axes, and $z$ is defined along the direction of $\mathbf{B}$. The hyperfine interaction with nuclear spins in the diamond lattice ($^{14}$N/$^{15}$N and $^{13}$C) is much smaller than the axial ZFS, including the largest hyperfine interaction with $^{13}$C (~1.1% natural abundance) in the first (~130 MHz) and third (~14 MHz) shells[4-6]. For the vast majority of NV⁻ centres (>96% of the NV⁻ centres without first shell $^{13}$C), the hyperfine interactions only lead to a very small perturbation and thus are not included in Eq. 1. The calculated energy levels of the NV⁻ centres of different orientations are shown in Fig. S2a.

## 1.2 The N$_S$ centre

The substitutional nitrogen defect (N$_S$) in diamond is also known as the P$_1$ centre (Fig. S1b). Like the NV⁻ centre, it also has four possible orientations due to the static Jahn-Teller distortion from T$_d$ to C$_{3v}$ symmetry[7]. In the spin-1/2 electronic ground state, the Hamiltonian of the N$_S$ centre is given by[8-10]:

$$H_{N_S} = \gamma_e \mathbf{B} \cdot \mathbf{S}^1 + \mathbf{S}^1 \cdot \mathbf{A}^1 \cdot \mathbf{I}^1 - \gamma_{^{14}N} \mathbf{B} \cdot \mathbf{I}^1 + Q^1 \cdot \left[ \left(I_z^1\right)^2 - \left(\mathbf{I}^1\right)^2 / 3 \right], \quad (S2)$$

where $\mathbf{S}^1$ is the electron spin of the N$_S$ centre ($S^1 = 1/2$), $\mathbf{I}^1$ is the $^{14}$N nuclear spin ($I^1 = 1$) with the nuclear gyromagnetic ratio $\gamma_{^{14}N}$, $\mathbf{A}^1$ is the hyperfine interaction tensor, given by its axial ($A_\parallel^1 = 113.982$ MHz) and transverse ($A_\perp^1 = 81.345$ MHz) components, and $Q^1 = -3.971$ MHz is the quadrupole coupling parameter. Here only $^{14}$N isotope is considered because the natural abundance of $^{14}$N and $^{15}$N is 99.634% and 0.366%, respectively. The calculated energy levels of N$_S$ centre of different orientations are shown in Fig. S2b.

## 1.3 The NV⁰ centre

The electronic structure of the NV⁰ centre is less well-understood than that of the NV⁻ or N$_S$ centre (Fig. S1c). The proposed spin-1/2 $^2E$ ground state has not been observed by EPR, probably because it is either obscured by other paramagnetic defects or broadened by the dynamic Jahn-Teller distortion[11]. A recent optically excited EPR experiment has assigned the observed transitions of a trigonal nitrogen-containing defect in diamond to the spin-3/2 $^4A_2$ excited state of the NV$^0$ centre, which should have relatively low energy and long lifetime. The Hamiltonian of the NV$^0$ centre is given by[11-13]:

$$H_{NV^0} = D^0 \cdot (S_z^0)^2 + \gamma_e \mathbf{B} \cdot \mathbf{S}^0 + \mathbf{S}^0 \cdot \mathbf{A}^0 \cdot \mathbf{I}^0 - \gamma_{14_N} \mathbf{B} \cdot \mathbf{I}^0 + Q^0 \cdot \left[(I_z^0)^2 - (\mathbf{I}^0)^2/3\right], \quad (S3)$$

Where $D^0$ is the axial zero-field splitting parameter with the value determined as $2 \times D^0 \approx 1685(5)$ MHz, $\mathbf{S}^0$ is the electron spin of the NV$^0$ centre ($S^0 = 3/2$), $\mathbf{I}^0$ is the $^{14}$N nuclear spin ($I^0 = 1$) of the NV$^0$ centre, $\mathbf{A}^0$ is the hyperfine interaction tensor given by its axial ($A_\parallel^0 = -35.7$ MHz) and transverse ($A_\perp^0 = -23.8$ MHz) components[11], $Q^0 = -4.654$ MHz is the quadrupole coupling parameter[13]. The spin Hamiltonian parameters were determined using >95% $^{15}$N enriched NV$^0$ and is used here with reasonably good approximation[11]. For simplicity, only the calculated energy levels of the NV$^0$ centre oriented parallel to $B_z$ are shown in Fig. S2c.

## 2   Average distance between neighbouring centres

The typical concentrations of the relevant paramagnetic centres (NV$^-$, N$_S$, and NV$^0$) in diamond are on the order of magnitude of ppm. In such dilute spin systems, the average distance between one NV$^-$ centre and its neighbouring centres (NV$^-$, N$_S$, or NV$^0$) can be estimated by setting the probability of finding no neighbouring centre within a distance $r$ of the NV$^-$ centre to 1/2[14]:

$$\exp\left[-(4\pi\rho r^3/3)\right] = 1/2, \quad (S4)$$

where $\rho$ is the concentration of the relevant defect species. The average distance $R$ is then simply given by $R = 0.55\rho^{-1/3}$, with $\rho(\text{NV}^-) \approx 2$ ppm and $\rho(\text{N}_S) \approx 50$ ppm as previously determined[15,16]. The average distance between two neighbouring NV$^-$ centres is thus

$R(\text{NV}^-) \approx 8$ nm and that between one NV⁻ centre and its neighbouring $N_S$ centre is $R(N_S) \approx 3$ nm. Such an average separation is a good estimation of effective detection distance based on the agreement between observed and calculated transition frequencies; the few anomalous NV⁻ centres with closer neighbours will experience significantly shifted transitions that are not evident in the measurements.

**Fig. S1**

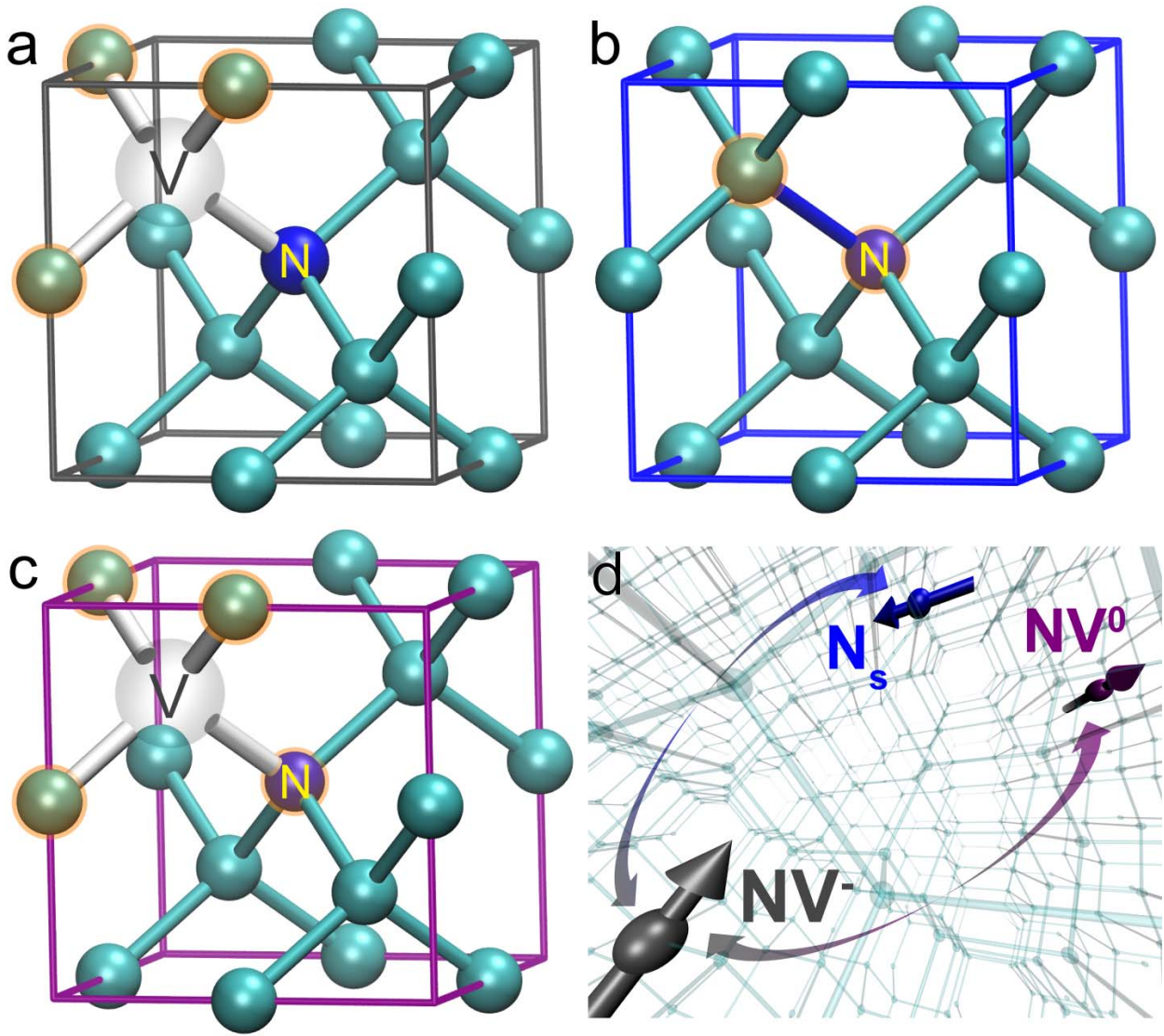

**Figure S1 | Structure of electron spin systems in diamond.** (a-c) The structures of the NV⁻ (a), $N_S$ (b), and $NV^0$ (c) centres. Transparent orange sphere indicates strong hyperfine coupling (>10 MHz) with the corresponding nuclear in the ground states of the NV⁻ and $N_S$ centres or in the $^4A_2$ excited state of the $NV^0$ center. (d) Illustration of the NV⁻ centre (grey spin symbol) in cross-relaxation with the nearest $N_S$ centre (blue spin symbol, approximately ~3 nm away) or $NV^0$ centre (purple spin symbol) in diamond.

**Figure S2**

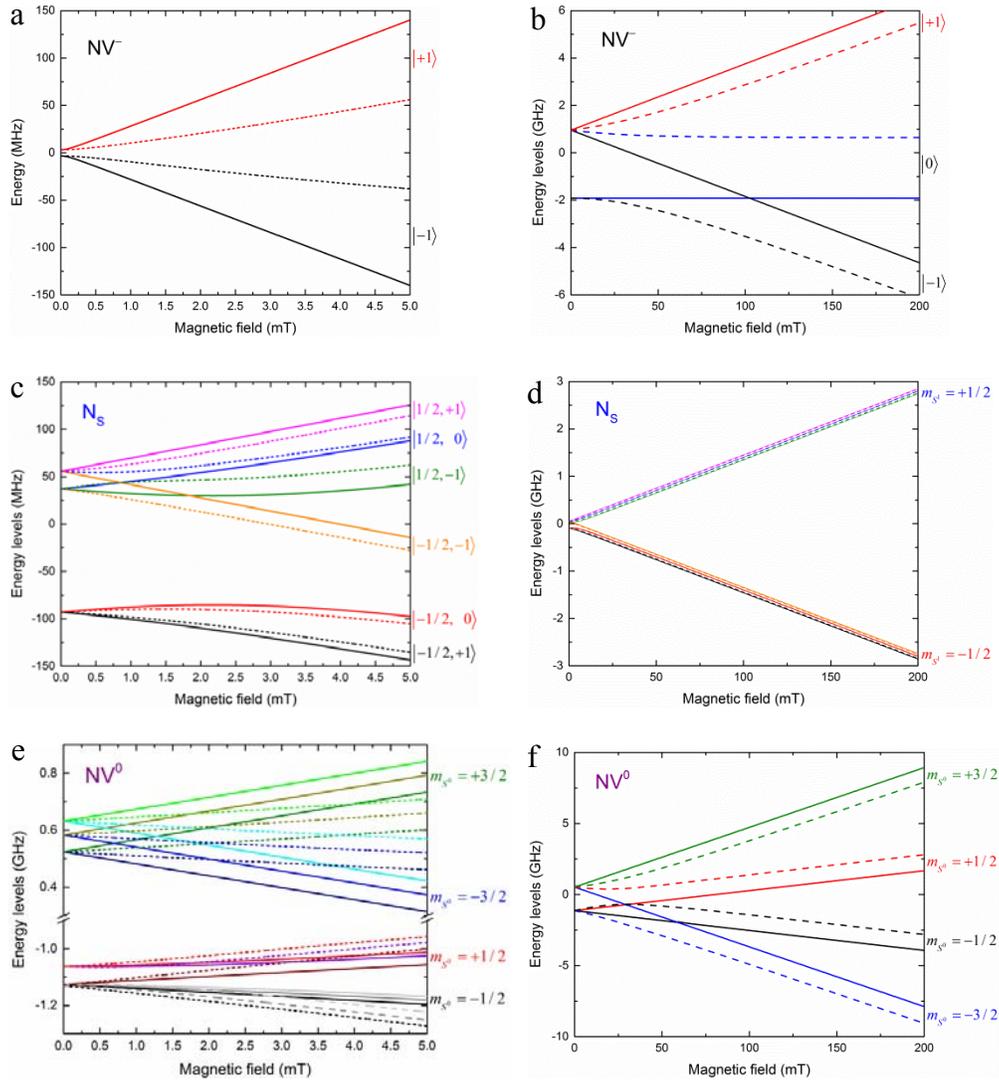

**Figure S2 | Calculated energy levels of the three spin systems in diamond.** The energy levels of the NV⁻ (a, b), N$_S$ (c, d), and NV⁰ (e, f) centres at weak magnetic fields (a, c, e) and between 0-200 mT (b, d, f) are calculated based on their respective Hamiltonians. The solid lines are the centres oriented parallel to $B_z$, and the dashed lines are centres ~109° relative to $B_z$. The energy levels of the N$_S$ centres are labelled by $|m_{S^1}, m_{I^1}\rangle$, where $m_S$ and $m_I$ are the electron and nuclear spin projections at the high field limit, respectively. The energy levels of the NV⁰ centres are labelled by the electron spin quantum number $m_{S^0}$.

**Figure S3**

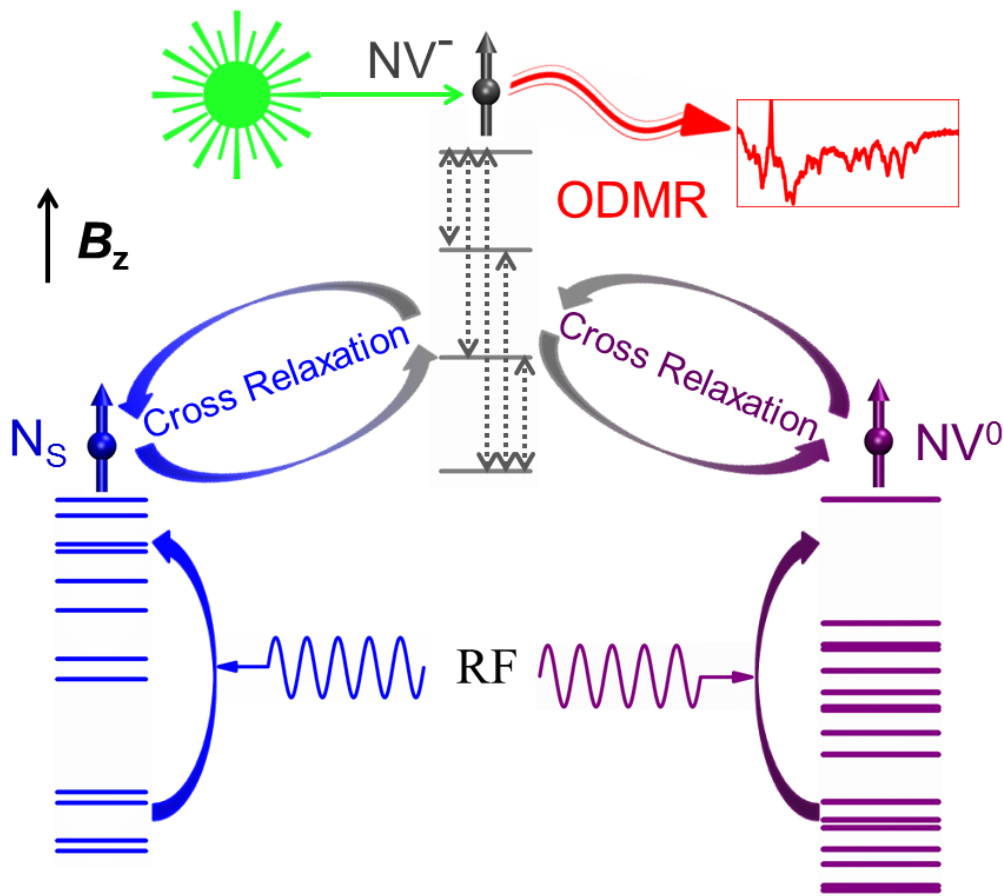

**Figure S3 | Schematic illustration of optically-detected cross-relaxation spectroscopy.** The cross relaxation occurs between the NV⁻ and $N_S$ centres, between the NV⁻ and $NV^0$ centres, and between magnetically inequivalent NV⁻ centres. When one of the $N_S$ or $NV^0$ transitions is saturated by RF radiation, the fluorescence intensity of the NV⁻ centres will decrease, resulting in the observed ODMR spectra.